\documentclass[11pt, a4paper]{article}
\pdfoutput=1
\usepackage{jheppub}
\usepackage{graphicx}
\usepackage{epsfig,amsmath,amsthm}
\usepackage{epstopdf}
\newcommand{\be}{\begin{equation}}
\newcommand{\ee}{\end{equation}}
\newcommand{\bea}{\begin{eqnarray}}
\newcommand{\eea}{\end{eqnarray}}
\def\bse{\begin{subequations}}
\def\ese{\end{subequations}}
 \newcommand{\IC}{\mathbb{C}}
 
\def\IZ{\relax\ifmmode\hbox{Z\kern-.4em Z}\else{Z\kern-.4em Z}\fi}

\newcommand{\non}{\nonumber \\}

\def\half{\frac{1}{2}} 
\def\del{{\partial}}

\def\etab{{\bar \eta}}

\def\cM{{\cal M}}  \def\cL{{\cal L}} \def\cH{{\cal H}}
 
  \def\eps{\epsilon}

  \def\tomega{\tilde{\omega}}

\newcommand{\revision}[1]{#1}

\def\presub{\vspace{.5cm} \noindent}

\def\bi{\begin{itemize}} \def\ei{\end{itemize}}

\def\({\left(} \def\){\right)}
\def\[{\left[} \def\]{\right]}
\def\<{\left<} \def\>{\right>}

\title{Natural dynamical reduction of the three-body problem}

\author{Barak Kol  \\
{\it Racah Institute of Physics, Hebrew University, Jerusalem 91904, Israel} \\
{\tt barak.kol@mail.huji.ac.il}
}

\abstract{The three-body problem is a fundamental long-standing open problem, with applications in all branches of physics, including astrophysics, nuclear physics and particle physics.  In general, conserved quantities allow to reduce the formulation of a mechanical problem to fewer degrees of freedom, a process known as dynamical reduction. However, extant reductions  are either non-general, or hide the problem's symmetry or include unexplained definitions. This paper presents a general and natural dynamical reduction, which avoids these issues. 

Any three-body configuration defines a triangle, and its orientation in space. Accordingly, we decompose the dynamical variables into the geometry (shape + size) and orientation of the triangle. The geometry variables are shown to describe the motion of an abstract point in a curved 3d space, subject to a potential-derived force and a magnetic-like force with a monopole charge. The orientation variables are shown to obey a dynamics analogous to the Euler equations for a rotating rigid body; only here the moments of inertia depend on the geometry variables, rather than being constant. 

The reduction rests on a novel symmetric solution to the center of mass constraint inspired by Lagrange's solution to the cubic. The formulation of the orientation variables is novel and rests on a partially known generalization of the Euler-Lagrange equations to non-coordinate velocities. Applications to global features, to the statistical solution, to special exact solutions and to economized simulations are presented. A generalization to the four-body problem is presented.}

\begin{document}

\maketitle

\begin{flushright}
To Joseph-Louis Lagrange \\
and George Lema\^itre
\end{flushright}

\section{Introduction}

The three-body problem involves the motion of three point masses influenced by mutual forces which depend on their separations.

While the two-body problem is integrable, the three-body problem is in general non-integrable and chaotic. Its solution has been a puzzle ever since it was studied by Newton in an attempt to understand the motion of the moon  \cite{Principia}. The system is known to be rather rich, and along the centuries its study deeply influenced several scientific fields including perturbation theory, chaos and topology.

It is interesting to seek a dynamical reduction of the system: a simplified formulation, where  the dynamical variables are chosen such that the equations of motion simplify. More specifically, the existence of conserved quantities allows to reduce the order of the system of differential equations.  While a dynamical reduction is often applied to integrable systems, it could also apply to a non-integrable system, in which case the reduced system remains non-integrable.

This work originated from the recent flux-based reduction of the statistical solution to the three-body problem in Newtonian gravity \cite{flux_based}, see also progress on the statistical solution through non-flux-based methods \cite{Stone_Leigh_2019,Ginat_Perets_2020}. The flux-based theory involves the (regularized) phase space volume as a function of the conserved charges. It is defined by a multi-dimensional integral, yet it was found that a certain change of variables enables analytic integration over numerous variables \cite{flux_based,sigma}. This suggested the possibility that certain dynamical reductions may exist, involving the new variables, thereby providing a deeper reason for the success of the analytic integrations.

Over the centuries, much \revision{magnificent} research was devoted to the dynamical reduction of the three-body problem (3BP). Lagrange (1772) \cite{Lagrange_1772} reduced the equations' order through replacing the position vectors by variables describing the triangle geometry and additional variables. Jacobi (1842) \cite{Jacobi_1842} showed how to perform one additional reduction of order, known as the elimination of the nodes. Levi-Civita (1915) \cite{Levi-Civita_1915} was the first to discuss the reduction at the level of the action, namely using a Lagrangian or a Hamiltonian, carrying out the reduction to the instantaneous three-body plane. Murnaghan (1936) \cite{Murnaghan_1936} performed a reduction of the planar 3BP at the level of the action, choosing side lengths as the dynamical variables.  However, the side length variables become singular for collinear configurations. Van Kampen and Wintner (1937) \cite{van_Kampen_Wintner_1937} generalized \cite{Murnaghan_1936} to 3d by adding the average azimuthal angle as a fourth coordinate.   Lema\^itre (1952, \revision{1964}) \cite{Lemaitre_1952,Lemaitre_1964} introduced a different set of symmetric coordinates, which remain smooth in the collinear limit, and presented a Hamiltonian for the general, non-planar case. \revision{These coordinates were given by a geometric construction and are related to the principal axes of inertia of the configuration}. More recently,  Moeckel and Montgomery (2013) \cite{Moeckel_Mont_2013} introduced the term shape sphere, for a set of symmetric coordinates for the planar 3BP, see also \cite{Montgomery2014,Montgomery2019}. These coordinates are essentially equivalent to those introduced by Lema\^itre, though it is not immediately apparent. In addition,  \cite{Moeckel_Mont_2013} recast the analysis into modern mathematical terminology. See appendix \ref{app:quotes} for selected quotes out of the above-mentioned papers.
 
The quantum version of the three-body problem was studied in the context of the Helium atom (nucleus + 2 electrons = 3 bodies). Interestingly, some works on the quantum version reached new definitions of the dynamical variables and new forms for the quantum Hamiltonian, which, in hindsight, are related with novel formulations of  the non-quantum problem. Hylleraas (1929) \cite{Hylleraas_1929} presented a Hamiltonian in terms of side lengths for the $\vec{J}=0$ sector and furthermore setting two electron masses to be equal and the nucleus mass to be infinite. Gronwall (1932) \cite{Gronwall_1932} introduced 3d variables that made the kinetic term conformally Euclidean. Fock (1954) \cite{Fock_1954} introduced 4d variables that reduce to Gronwall's variables after imposing the reduction to $\vec{J}=0$. Finally, Tkachenko (1978)  \cite{Tkachenko_1978} presented a Hamiltonian for non-zero $\vec{J}$.
 
\revision{In mathematics, the three-problem inspired considerable development, and one can mention the following contemporary works. \cite{McGehee_1974}  introduced the triple collision manifold in the context of linear motion, and \cite{Waldvogel_1982} generalized it to planar motion. \cite{Albouy_Chenciner_1998} generalized the reduction and the determination of the shape preserving time evolutions of the $n$-body system to arbitrary space dimensions.} Lastly, the 3BP is the setup for one of Smale's problems for 21st century mathematics (no. 6) \cite{Smale}.

The study of the three-body problem goes beyond the above-mentioned extensive work and includes diverse work within general physics, quantum physics, computer simulations and astrophysics. Forces other than Newtonian gravity were considered, and a sample of such work includes Coulomb forces \cite{Liverts_Barnea_2013} and harmonic forces \cite{Principia} prop. 64, \cite{Saporta_Katz-Efrati_2020}.  
 
Despite this venerable history, certain features of the reduction are still lacking. First, a choice of coordinates that satisfy the center of mass constraint appears to violate the permutation symmetry (see section \ref{subsec:translations}), which is unnatural. Secondly and relatedly, the shape sphere variables display somewhat surprising and undeserved properties (see last paragraph of section \ref{subsec:discussion}), which suggest that a more direct and natural route to their introduction might exist, which would make them clearer. Thirdly and lastly, extant formulations do not incorporate the angular momentum components as canonical variables, which would be natural given their central standing as conserved quantities.

In this paper, the reduction process is described in section \ref{sec:reduction}. It starts by introducing a solution to the center of mass constraints, which transforms nicely under permutations, and hence is natural. It involves complex numbers in a novel way, and it is inspired by Lagrange's solution to the cubic. 
 Next, the orientation of the plane defined by the three bodies is separated from motion within the plane simply by using a rotating body frame. We proceed to the separation of rotations within the plane from the geometry of the triangle formed by the three bodies. This requires to define certain spherical coordinates on $\IC^2$ and to handle correctly one of the angles, resulting in a famous quotient of it. Finally, we Legendre transform from angular velocities to angular momenta and obtain their Poisson brackets from a general theory for non-coordinate velocities and their conjugate momenta. This leads us to our final result, \eqref{L_J}. We end the section with an illustration of the resulting formulation and a discussion of it.

In section \ref{sec:appl}, we present several applications of the formulation. We define Hill-like regions in geometry space. We discuss the total phase volume of the three-body system, which is an ingredient of the statistical theory of \cite{flux_based}. We provide a novel derivation of the uniformly rotating planar solution. Finally, we mention implications for economizing simulations.

This work has interesting implications to the four-body problem, and a sketchy generalization appears in section \ref{sec:4body}. Finally, section \ref{sec:epilogue} is a brief epilogue.

Further appendices provide background material on Lagrange's solution to the cubic, on bi-complex numbers and the isotropic oscillator, and finally, on the equations of motion in terms of non-coordinate velocities and the rotating rigid body. 

\section{Reduction}
\label{sec:reduction}

\noindent {\bf Setup}. Consider three point masses, $m_1, m_2, m_3$. Let us take the bodies' position vectors, $\vec{r}_a, ~a=1,2,3$, to be our initial dynamical variables. The kinetic energy is given by \be
	T :=  \sum_{a=1}^3 \half\, m_a\, \dot{\vec{r}}_a^{~2}  \label{def:T}
\ee
	For concreteness, take the potential to be given by the Newtonian gravitational potential \be
	V := - \frac{G\, m_1\, m_2}{r_{12}} -  \frac{G\, m_1\, m_3}{r_{13}} - \frac{G\, m_2\, m_3}{r_{23}} 
	\label{def:V}
\ee
where $G$ is Newton's gravitational constant, and $r_{ab}=\left| \vec{r}_a - \vec{r}_b \right|$, are the inter-body distances. We note that the whole reduction procedure is independent of the choice of $V$ as long as it depends on $r_{ab}$ only, namely $V=V(\{ r_{ab} \} )$. Finally, the system can be  defined by the Lagrangian \be
	\cL\( \{\vec{r}_a,\, \dot{\vec{r}}_a \}_{a=1}^3 \) := T - V \label{def:L_r} ~.
	 \ee  

\subsection{Translations and the complex position vector $\vec{w}$}
\label{subsec:translations}

The system is invariant under translations and hence the total linear momentum is conserved, the center of mass moves in uniform motion, and henceforth, we work in the center of mass frame \be
 0=\vec{r}_{CM} := \frac{1}{M} \sum_a m_a \vec{r}_a
 \ee
where $M := m_1 + m_2 + m_3$ is the total mass.

The center of mass constraint reduces the configuration space, reducing the number of degrees of freedom from 9 to 6. This requires a choice of coordinates on the reduced configuration space. We shall see that this choice creates a tension with the bodies' permutation symmetry, which will affect the whole reduction process, and hence, we shall describe it in some detail. 

The coordinates are chosen to be translation-invariant vectors: vectors, in order to preserve the transformation properties under rotations, and translation-invariant, in order to decouple them from the center of mass coordinates in the expression for the kinetic energy.

In the two-body problem, one chooses the relative coordinate to be \be
	\vec{r}_1-\vec{r}_2
\ee
It is a translation-invariant vector, and under a $1 \leftrightarrow 2$ exchange, it changes only by a sign.

In the three-body problem, there are three relative vectors \be 
	\vec{r}_{12}:=\vec{r}_1-\vec{r}_2, \qquad \vec{r}_{23}:=\vec{r}_2-\vec{r}_3, \qquad \vec{r}_{31}:=\vec{r}_3-\vec{r}_1
\ee
They satisfy the constraint $0= \vec{r}_{12} + \vec{r}_{23} + \vec{r}_{31}$. 

Any pair of linear combinations of the relative vectors can serve as coordinates in the center of mass frame. There are two popular choices for that. The first set consists of the planetary coordinates \be
 	\vec{r}_{13}, \qquad \vec{r}_{23}
	\ee
They are useful for the case where one of the masses, say $m_3$, is much heavier than the other two: $m_1,m_2 \ll m_3$. The second set consists of the lunar coordinates \be
	\vec{r}_{12}, \qquad \frac{1}{m_1+m_2} \( m_1\, \vec{r}_1 + m_2\, \vec{r}_2 \) - \vec{r}_3 
\ee
These are useful for the hierarchical case where the magnitude of the first vector is much small than the second, namely bodies 1 and 2 are much closer to each other, relative to their distance from the third body. These coordinates are commonly called Jacobi coordinates, because they appeared in \cite{Jacobi_1842}, but they are a rather natural choice, and they appeared already in \cite{Principia,Lagrange_1772}. 

Both coordinate choices break the symmetry between the bodies. Hence, they are not optimal for the non-hierarchical case where all distances and all masses are comparable. 

Can we define natural, namely symmetric, coordinates in the center of mass frame? A solution is suggested by Lagrange's solution to the cubic equation. Lagrange realized that the key to a solution of a polynomial equation is to study expressions in terms of the roots that display a certain measure of symmetry among them. For more see appendix \ref{app:cubic}. Inspired by this, we define \be
 \vec{w} := \vec{r}_1 + \eta \, \vec{r}_2 +\etab\, \vec{r}_3
 \label{def:w}
 \ee
 where $\eta$ is the cubic root of unity, namely $\eta=\exp (2 \pi j /3)$, and we denote here the imaginary unit by $j$, $j^2=-1$, since $i$ will have a different use below. 
 
The definition \eqref{def:w} is one of the central results of this paper. Let us discuss it. By construction $\vec{w}$ is a complex vector. Naturally, it is composed of 2 real vectors, and so it has the correct number of components to serve as coordinates in the center of mass frame. It is translation-invariant as a result of the identity  $1+ \eta + \etab=0$. Under a cyclic rotation it is merely multiplied by a $2 \pi/3$ phase, while the exchange $2 \leftrightarrow 3$  transforms it into its complex conjugate. \emph{These simple transformations under permutations render $\vec{w}$ a natural coordinate}. 

\presub {\bf Transformed Lagrangian} Given $\vec{w}$, the relative positions can be expressed by \be
 \vec{r}_{23}=\frac{1}{j \sqrt{3}} \( \vec{w} - \vec{\bar w} \) 
\label{r_23_w}
 \ee
and similarly, replacing $w \to \eta w$ in this expression gives $\vec{r}_{12}$, and $w \to \etab w$ gives $\vec{r}_{31}$.

The kinetic energy in the center of mass frame is given by \bea
	T_{CM} &:=& T -  \frac{M}{2}\, {\dot{\vec{r}}_{CM}}^2 = \frac{1}{2 M} \( m_2\, m_3\,  {\dot{\vec{r}}_{23}}^2 + cyc. \) = \non
	&=& \frac{1}{6 M} \( m_2\, m_3\, \lVert \frac{d}{j\, dt}(\vec{w}-\vec{\bar w}) \rVert^2 + cyc. \)
\label{T_CM}
\eea
where the cyclic transformation denotes now $w \to \etab w$, in addition to $1 \to 2 \to 3 \to 1$.

\subsection{Rotations and the orientation--geometry decomposition}

One can naturally divide the system's coordinates into orientation coordinates and coordinates that describe the configuration up to rotations, namely geometry coordinates. At any given moment, the positions of the three bodies define an instantaneous plane \revision{(apart for collinear configurations, which are codimension 2 in configuration space and hence can be ignored for now -- for more see the end of Section \ref{subsec:geom}).}

We define a rotating body frame whose $z$ axis is normal to the instantaneous plane \revision{(either choice of the positive direction would do, namely, this is a $\mathbb{Z}_2$ gauge, and once chosen it remains determined throughout time evolution as long as collinear configurations are not reached)}. 2d vectors $\vec{\rho}_a$ specify the positions within the rotating $(x,y)$ plane. As usual,  inertial frame velocities are given by \be
 \left. \frac{d}{dt} \right|_{inert} \vec{\rho}_{ab} = \dot{\vec{\rho}}_{ab} + \vec{\tomega} \times \vec{\rho}_{ab}
 \label{inertial_vel}
 \ee
 where $\vec{\tomega}$ is the angular velocity vector that describes the rotation of the body frame, and $\omega$ is reserved for later use. 
 
Substituting, the kinetic energy \eqref{T_CM} becomes \be
  T = \half I_{ij}\, \tomega^i\, \tomega^j + \vec{L}_w \cdot \vec{\tomega} +  T_w 
 \label{T_tomega}
 \ee
where $I_{ij}, ~ i,j=1,2,3$ denotes the inertia tensor in the rotating (and center of mass) frame \be
 I_{ij} = \frac{1}{M} \[ m_2\, m_3\, \(  \delta_{ij} \vec{\rho}_{23}^{\;2} - \rho^i_{23} \rho^j_{23} \) + cyc. \] ~;
\label{def:I_ij} 
 \ee
$\vec{L}_w$ denotes the angular momentum due to the $\vec{\rho}_a$ motion \bea
  \vec{L}_w &=& L_w \hat{z} \non
 L_w &=& \frac{1}{M} \( m_2\, m_3\,  \vec{\rho}_{23} \wedge \dot{\vec{\rho}}_{23}  + cyc. \) ~, 
 \label{def:L_w}
   \eea
  where the wedge product between any 2d vectors $\vec{u},\vec{v}$ is defined by $\vec{u} \wedge \vec{v} := (\vec{u} \times \vec{v}) \cdot \hat{z}$; and finally, the kinetic energy due to the $\vec{\rho}_a$ motion \be
  T_w := \frac{1}{M} \( m_2\, m_3\,  \dot{\vec{\rho}}_{23}^{\; 2} + cyc. \)  ~.
\label{def:T_w}
  \ee
  
We note several properties of $I_{ij}$. Since the mass distribution is confined to the $z=0$ plane, one has \bea
 0	 &=& I_{13}= I_{23} \non
  I 	&:=& I_{33} = I_{11} + I_{22}
 \eea
In addition, since the mass distribution consists of only 3 point masses, one has \be 
 I_{11}\, I_{22} - \(I_{12}\)^2 = \frac{m_1\, m_2\, m_3}{M}\, \Delta^2
 \label{det_I}
 \ee
 where \be \Delta := \det \[ x_1, y_1, 1; x_2, y_2, 1; x_3, y_3,1 \]
 \label{def:Delta}
 \ee  
 is twice the (signed) area of the triangle formed by the bodies. In other words, the $2*2$ determinant of $I_{\alpha \beta}, ~\alpha, \beta=1,2$ factors into a part that depends only on positions, and a part which depends only on the masses. 

\subsection{Rotations within the plane and geometry space}

The orientation--geometry decomposition is not done yet. To see that, let us count the number of generalized coordinates. We know that the 3d problem has 6 degrees of freedom (d.o.f) in the center of mass frame (CM). The orientation is specified by 3 angles, such as the Euler angles, and the $\vec{\rho}_a$ at CM (or equivalently the associated 2d $\vec{w}$) provide 4 other d.o.f., so altogether we have 7 generalized coordinates. This means that these coordinates are redundant by a single degree of freedom.

Rotations within the $(x,y)$ plane are the origin of this redundancy: since a rotation of the body frame around the $z$-axis is equivalent to a rotation of the 2d vectors, the configuration depends on the two associated angles only through their sum. 

\presub {\bf Bi-complex $w$}. In order to account for plane rotations, it is useful to complexify the plane, so that rotations would be represented by a phase multiplication. By complexification one means that any 2 vector $\vec{\rho}$ is mapped onto a complex number  $\vec{\rho} \to \rho := \vec{\rho} \cdot \( \hat{x} + i\, \hat{y} \)$. Similarly, the 2d $j$-complex vector $\vec{w}$, defined in \eqref{def:w}, is mapped onto a so-called bi-complex number \be
	\vec{w} \to w = \vec{w} \cdot \( \hat{x} + i\, \hat{y} \)
	\label{def:w-bi}
\ee 
which is of the form $w = a + b\, i + c\, j + d \, i j$. The imaginary unit $i$ represents a quarter rotation in the $(x,y)$ plane. Hence, $i,j$ are commuting imaginary units, namely $i^2=j^2=-1$ and $ i\, j = j\, i$.  For more on the algebraic structure of bi-complex numbers, and their application to the isotropic oscillator of mechanics, see appendix \ref{app:bicomplex}.  

Let us introduce a natural basis for the $w$ space. Evaluating the $w$ variable \eqref{def:w-bi} for a right-handed and a left-handed equilateral triangles of unit sides motivates the definitions \be
 e_R := \frac{\sqrt{3}}{2} \( 1+ i\, j\) \qquad e_L :=  \frac{\sqrt{3}}{2} \( 1- i\, j\) ~,
 \ee
where for concreteness, the equilateral triangles are chosen to be oriented such that the height from edge 2---3 to vertex 1 is pointed along $+\hat{x}$ within the $(x,y)$ plane. \revision{Algebraically, $e_R, \, e_L$ are characterized as being zero divisors of the bi-complex ring (see Appendix \ref{app:bicomplex}), and thereby are natural.}

\presub {\bf Spherical coordinates for $w$}. Considering $w$ space as a 2d complex vector space over complex numbers involving $i$, namely $\IC[i]^2$, one can expand a general state $w$ in the $e_R,\, e_L$ basis as follows
\be
	w = r\, e^{i \psi_0}\, \[ e^{i \phi/2}\, \cos \frac{\theta}{2}\, e_R + e^{-i \phi/2}\, \sin \frac{\theta}{2}\, e_L \] ~.
\label{def:spher}
\ee
The definition of $\psi_0$ can be changed into $\psi = \psi_0 + \chi(\theta, \phi)$, which is a gauge transformation. The variable $r$ sets the overall triangle scale, $\psi \sim \psi + 2\pi$ is an overall rotation angle, $\theta \in [0,\pi]$ determines the relative magnitudes of the right and left components, and finally $\phi \sim \phi + 2 \pi$ is the relative phase of the right and left components. Altogether $r,\theta,\phi,\psi$ are spherical coordinates for $w \in \IC[i]^2$. 

We define $G$ to be the quotient space of planar three-body configurations, up to rotations, where $G$ stands for geometry. In other words, $G$ is the space of equivalence classes of congruent triangles. We can write \be
 G = \IC[i]^2/U(1)
\label{quotient}
\ee
where the $U(1)$ acts by overall phase rotations. We shall later see how to parameterize $G$ in terms of quadratics in $w$. In coordinates, $G$ is given by the $w$ variable up to $\psi$-shifts, and it is parameterized by $r,\theta,\phi$. $G$ includes information on both size and shape: the $r$ coordinate determines the triangle size, while the $\theta, \phi$ coordinates parameterize a sphere that describes triangles up to similarity, and is known as the \emph{shape sphere}.
Coordinates equivalent to $r,\theta,\phi$ appeared in \cite{Lemaitre_1952} in Eqs. (1,2,14), \revision{where they were found through a trial and error process: \cite{Lemaitre_1952} started with the equal mass case, used its increased symmetry, and then generalized to unequal masses, while \cite{Lemaitre_1964} started from equal moments of inertia and then generalized.} The term ``shape sphere'' appeared  in \cite{Moeckel_Mont_2013}, \revision{see also \cite{Easton_1971,Saari_1984,Moeckel_1988}.}

Substituting \eqref{def:spher} into \eqref{r_23_w}, the relative position vector $r_{23}$ becomes \be
	 \rho_{23} = i\, r\, e^{i \psi_0} \[ e^{i \phi/2}\, \cos \frac{\theta}{2}\, - e^{-i \phi/2}\, \sin \frac{\theta}{2}\, \] ~.
\ee
 In particular,  \be
 r_{23}^{~2} \equiv \left|  \rho_{23} \right|^2 = r^2\, \( 1 - \sin \theta\, \cos \phi \) ~.
 \label{r_23-G}
 \ee
Similarly, $r_{31}, r_{12}$ are given by the same expression after the substitutions $\phi \to \phi - 2 \pi/3$, and $\phi \to \phi + 2 \pi/3$, respectively. This means that expressions are invariant under a cyclic permutation combined with a third of a revolution in $\phi$.

$\psi_0$, defined in \eqref{def:spher}, is symmetric between the $R$ and $L$ hemispheres, $0 \le \theta \le \pi/2$ and $\pi/2 \le \theta \le \pi$, respectively. However, $\psi_0$ is singular for both $R$ and $L$ poles. A gauge that is regular at the $R$ pole is given by \be
	\psi_+ := \psi_0 + \phi/2 
\label{def:psi+}
\ee

\presub {\bf Kinetic energy}. In spherical coordinates, and using the $\psi_+$ gauge \eqref{def:psi+}, $T_w$ \eqref{def:T_w} becomes \be
 T_w = \half\, I\, \( \dot{\psi}_+ +\frac{L_G}{I} \)^2 + T_G
\label{T_w}
 \ee
where the largest principal moment of inertia (for rotations within the plane) \eqref{def:I_ij}, is given by \be
 I := I_{33} = \frac{r^2}{M} \[ m_2\, m_3 \(1 - \sin \theta\, \cos \phi\) + cyc. \] ~.
\label{I-G}
\ee
 $L_G$ denotes the angular momentum in the geometry space $G$  and it is given by  \be
 L_G = -\frac{r^2}{2\, M} \[ m_2\, m_3 \( \(1 - \cos \theta - \sin \theta \, \cos \phi \) \dot{\phi} - \sin \phi \, \dot{\theta} \) + cyc.  \] ~
\label{def:L_G}
 \ee
 where $cyc.$ denotes $1 \to 2 \to 3 \to 1$ together with $\phi \to \phi - 2 \pi/3$, just like the comment after \eqref{r_23-G}. $L_G$ is related to $L_w$, defined in \eqref{def:L_w}, by \be
 L_w=I\,\dot{\psi}_+ + L_G
 \label{L_w-L_G}
 \ee
 $T_G$ denotes the kinetic energy in $G$ space, and is given by \be
 	T_G :=\frac{1}{8\, I}\, \dot{I}^2  + \frac{3\, M_3}{8\, M} \frac{r^4}{I} \(\dot{\theta}^2 + \sin^2 \theta\, \dot{\phi}^2 \) ~.
\label{def:T_G}
\ee
Finally, $M, M_2, M_3$ denote the elementary symmetric functions of the masses  \bea
 M &:=& m_1 + m_2 + m_3 \non
 M_2 &:=& m_2 \, m_3  + m_3 \, m_1  + m_1 \, m_2 \non
 M_3 &:=& m_1\, m_2\, m_3 ~.
\eea 

The derivation proceeds by projecting $\dot{\rho}_{23}$ along the radial and the tangential directions with respect to $\rho_{23}$, performed in the $r,\theta,\phi$ coordinates, and similarly for the other relative velocities. It is rather lengthy and non-illuminating and I suspect that a better one can be found, perhaps along the lines of \cite{Moeckel_Mont_2013}. For these reasons, it will not be included here. 

The kinetic energy \eqref{T_w} is in a form of a dimensional reduction (also known as Kaluza-Klein reduction) over the coordinate $\psi$. Therefore,  the expression for $L_G$ depends on the choice of gauge for $\psi$, while the kinetic metric $T_G$ is gauge-independent, and represents the metric on the quotient space \eqref{quotient}. The expression for $T_G$ appeared essentially in \revision{Eq. (4.3.13) of \cite{Montgomery_2002}, which studied the $J^2=0$ sector,} and in \cite{Moeckel_Mont_2013}.

\presub {\bf Fixing the $\psi$ gauge}. Let us return to the issue of coordinate redundancy discussed at the beginning of this subsection. This redundancy can be removed by fixing a gauge for the $\psi$ coordinate, and we choose to set \be
	\psi=0 ~.
\ee
This can always be achieved through a choice of the $x$---$y$ body axes, which in turn fixes the value of $\tomega_z$. Clearly, the effect of this gauge is directly related to the choice of gauge for $\psi$.

Relatedly, the kinetic energy depends on $\tomega_z,\, \dot{\psi}$ only through their sum. Indeed, the relevant terms in $T$ are (\ref{T_tomega},\ref{T_w})
\bea
 T &\supset&   \half I \( \tomega_3 \)^2 + L_w \, \tomega_3 + \half I  \( \dot{\psi}+\frac{L_G}{I}\)^2 = \non
 	&=& \half I  \( \tomega_3 + \dot{\psi}+\frac{L_G}{I}\)^2 = \half I  \( \omega_3 +\frac{L_G}{I}\)^2
\eea
where in passing to the second line, we have used \eqref{L_w-L_G}, and in the last equality we have used $\vec{\omega}$ defined by \be
	\vec{\omega} = \vec{\tomega} + \dot{\psi}\, \hat{z} ~.
\ee

Altogether, the expression for the kinetic energy in the center of mass frame becomes \be
	T\[\vec{\omega},r,\theta,\phi\] = T_G + \half I  \( \omega_3 +\frac{L_G}{I}\)^2 + \half I_{\alpha \beta}\,  \omega^\alpha \,  \omega^\beta
\label{T_in_G_var}
\ee
where $T_G,\, I,\, L_G,\, I_{\alpha\beta} ~\alpha,\beta=1,2$ depend on the $G$-space variables and were defined in (\ref{def:T_G},\ref{I-G},\ref{def:L_G},\ref{def:I_ij}) respectively. 

\revision{
An alternative gauge choice, suggested by \cite{Lemaitre_1964}, is to employ the principal axes of inertia also within the instantaneous plane. It has the advantage of being natural and of reducing the rotation equations \eqref{J_eq} to the familiar Euler equations.
}

\presub {\bf Interpretation as a 1d $SO(2)$ gauge theory}. The general formulation in a rotating frame \eqref{inertial_vel} can be considered to be an $SO(3)$ gauge theory in $0+1$ dimensions. In fact, this relation is the definition of the covariant derivative D/Dt, where $\vec{\tomega}$ is the gauge field in the adjoint representation of $so(3)$, namely ${\mathbf 3}$, and the charged fields are the position vectors $\vec{r}_a$, which transform in the same representation. Finite gauge transformations correspond to a frame redefinition through an $SO(3)$ rotation matrix.

In the three-body problem, we partially fix this gauge through $z_a=0 ~a=1,2,3$. We are left with a 1d $SO(2) \simeq U(1)$ gauge theory, the residue of the original $SO(3)$ gauge theory. The gauge field is $\tomega_3$, and the charged fields are the 2d position vectors $\rho_a$.

After transforming to spherical coordinates, $\psi$ remains the only charged field, such that $\dot{\psi}+\tomega_3$ is gauge invariant. Changing variables into $\omega_3$ fully eliminates the gauge field (this is possible for 1d gauge theories).
 
\subsection{Geometry of geometry space}
\label{subsec:geom}

Let us study further the geometry of geometry space, namely three-body configurations up to rotations \eqref{quotient}.

\presub {\bf Invariants}. So far, geometry space was described by the coordinates $r,\theta,\phi$. Alternatively, we can employ the invariants of the quotient. This approach will clarify the geometry at the origin of geometry space. 

We define the quadratic invariants \be
 Q_s := \[ \begin{array}{cc} w^*_1 & w^*_2 \end{array} \] \;
  \tau_s \;
 \[  \begin{array}{c} w_1 \\ w_2 \end{array} \] \qquad s=0,1,2,3,4
 \ee 
 where $w_1, w_2 \in \IC[i] $ denote the $j$-real and imaginary parts of the bi-complex $w$, up to normalization, defined as follows \be
	w = \sqrt{\frac{3}{2}} \( w_1 + j\, w_2 \) 
	\ee
 and the Pauli-like matrices  $\tau_r$ are given by \be
 \tau_0 = \[ \begin{array}{cc} 1 & 0 \\ 0 & 1 \end{array} \] ,~
 \tau_1 = \[ \begin{array}{cc} 1 & 0 \\ 0 & -1 \end{array} \] ,~
 \tau_2 = \half \[ \begin{array}{cc} -1 & -\sqrt{3} \\ -\sqrt{3}  & 1 \end{array} \] ,~
 \tau_3 =  \half \[ \begin{array}{cc} -1 & \sqrt{3} \\ \sqrt{3} & 1 \end{array} \] ,~
 \tau_4 = \[ \begin{array}{cc} 0 & -i \\ i & 0 \end{array} \] 
 \ee 
Since $Q_s$ are of the form $\sim w^*\, w$, they are manifestly invariant under the rotations $w \to \exp (i \psi)\, w$. 

The quadratic invariants $Q_0, \dots, Q_4$ are not all independent, but rather satisfy the relations \bse \begin{align}
 0 &= Q_1 + Q_2 + Q_3 \label{Qrela} \\
 Q_0^2 &= \frac{2}{3} \( Q_1^2 + Q_2^2 + Q_3^2 \)  + Q_4^2 \label{Qrelb}
 \end{align}
 \label{Qrel}
 \ese
Geometrically, relation \eqref{Qrela} means that $Q_1, Q_2, Q_3$ describe a flat 2d space, and relation \eqref{Qrelb} means that the quotient is a (light) cone in a 3+1 space. The cone singularity at the origin corresponds to a triple collision configuration. 

We note that the $Q_0,\dots Q_4$ variables are closely related to the Stokes parameters $S_0, \dots, S_3$ \cite{Stokes1852}, which are used to describe states of polarization of light and are reviewed in appendix \ref{app:bicomplex}. More precisely, $Q_0,\, Q_4$ are identical to the $S_0,\, S_3$ while $Q_1,Q_2,Q_3$ up to the relation \eqref{Qrela} are analogous to $S_1,S_2$. However, the $Q$ variables are distinguished by being compatible with a symmetry of order 3. 

In terms of the $r,\theta,\phi$ coordinates, substitution of \eqref{def:spher} shows that the invariants are given by \bse
\begin{align}
 Q_0 &= r^2		 \label{Q0_spher} \\
 Q_a &= r^2 \sin \theta \, \cos \phi_a, ~ a=1,2,3	\label{Qa_spher} \\
 Q_4 &= r^2\, \cos \theta \label{Q4_spher}
\end{align} \label{Q_spher} \ese
where \be
\phi_a := \phi-(a-1) 2 \pi/3 ~.
\ee
In this form, the invariants are seen to be closely related to Cartesian coordinates $\vec{G}$ associated with the spherical coordinates $r^2,\theta,\phi$, namely $G_1=r^2 \sin \theta \, \cos \phi, ~G_2=r^2 \sin \theta \, \sin \phi,$ and $G_3=r^2 \cos \theta$. The three components of $\vec{G}$ are independent coordinates, which solve the relations on the $Q$ variables \eqref{Qrel}. Similarly, we define the vector $\vec{g}$ to be the Cartesian coordinates associated with  $r, \theta, \phi$ ($r$ instead of  $r^2$).

Using \eqref{r_23-G}, the triangle edge lengths can now be expressed as \be
 r_{23}^{~2} = Q_0 - Q_1 ~,
 \ee
 and similarly for $r_{31}^{~2}, ~ r_{12}^{~2}$. Summing all three, and using (\ref{Qrela},\ref{Q0_spher})  we find \be
  r^2 = \frac{1}{3} \( r_{23}^{~2} + r_{31}^{~2} + r_{12}^{~2} \) ~.
  \ee
This means that the geometric interpretation of the radial coordinate $r$ is the root mean square of the side lengths.

We note that one could take a wider perspective, and rather than study the planar configuration of the bi-complex $w$ up to phase shifts, one could study the 3d complex position vector $\vec{w}$ \eqref{def:w} up to identification by 3d rotations, namely $\vec{w}/SO(3)$. The quadratic invariants $Q_0, \dots Q_3$ can be expressed in terms of scalar products among $\vec{w}, \, \vec{\bar{w}}$, and hence are invariants from the 3d perspective. On the other hand, $Q_4$ cannot be described in this way, and hence while it is an invariant of $\IC^2/U(1)$, only $Q_4^{~2}$ is an invariant of $\vec{w}/SO(3)$.

In addition, we note that $Q_4$ is proportional to the triangle area, more precisely \be
	Q_4 = \frac{2}{\sqrt{3}}\, \Delta
\label{Q_4-Delta}
\ee
where $\Delta$ is twice the (signed) triangle area, and was defined in \eqref{def:Delta}. This can be shown by following the definition of $Q_4$ and using translation invariance to set $\vec{\rho}_1=0$. \revision{Interestingly, \cite{Montgomery_2002} has shown that \eqref{Qrelb} is equivalent to Heron's formula for a triangle's area.}

\presub {\bf Shape sphere}. Geometry up to size is known as shape. For this reason, the unit sphere in geometry space, namely $r=1$, is known as shape sphere.

For triangles, shape is the same as classification up to similarity, which is well known to be classified by the values of the three angles which are constrained to sum to $\pi$. This means that shape space should be a 2d surface. As we have seen, it turns out that this surface has the topology of sphere. More precisely, the shape sphere is the space of shapes of triangles with labeled vertices.  

\begin{figure}
\centering \noindent
\includegraphics[width=8cm]{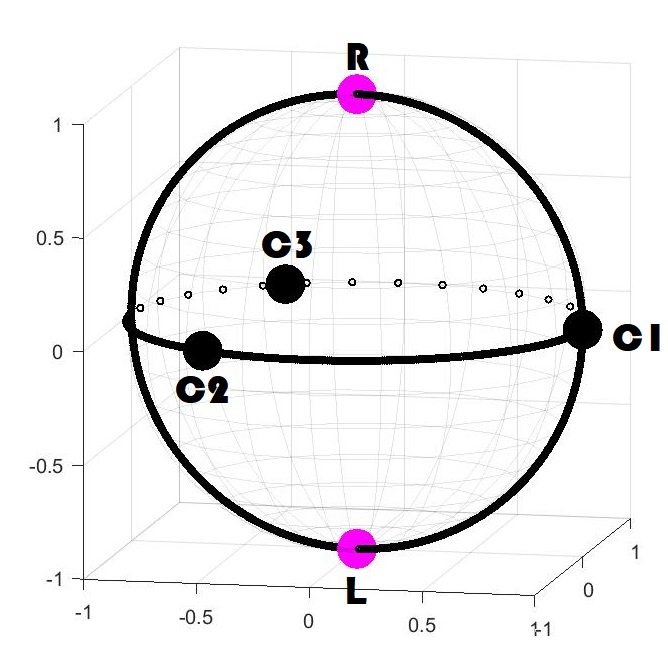} 
\caption[]{The shape sphere describes the space of all possible labeled triangles in a plane up to similarity. See text for the description of various locations on it. The vertical direction represents $Q_4$, while the horizontal directions represent $Q_1, Q_2, Q_3$ constrained by \eqref{Qrela}.}
 \label{fig1:shape_sphere} \end{figure}

The shape sphere is shown in fig. \ref{fig1:shape_sphere}. The geometrical interpretation of various locations on the shape sphere is known \cite{Moeckel_Mont_2013} and will be repeated here for convenience. The vertical $Q_4$ coordinate is proportional to the triangle area, as stated in \eqref{Q_4-Delta}. Hence, the $Q_4=0$ equator corresponds to collinear configurations. Within the collinear equator, the point $Q_1=1$ implies $w_2=0$ and hence it corresponds to the collision of the 2,3 vertices and it is denoted by $C1$. Similarly for $C2, C3$.  Going away from the equator, the $Q_4=1$ pole is associated with $w =1 + i\, j$, which corresponds to a right equilateral triangle, namely, such the 1,2,3 vertices are oriented in the positive mathematical direction (counter-clockwise). Similarly, the $Q_4=-1$ pole corresponds to a left equilateral triangle.

Just as the $Q$ invariants were noted above to be closely related to the Stokes parameters and the Pauli matrices (Pauli, 1927) \cite{Pauli1927}, the shape sphere is closely related to the Poincar\'{e} sphere, the sphere of polarizations that was introduced in Poincar\'{e}'s lectures on optics \cite{Poincare1892}. These notions are also closely related to the Hopf fibration (Hopf, 1931) \cite{Hopf1931} and the Bloch sphere (Bloch 1946) \cite{Bloch1946}. The Pauli matrices are used to describe quantum operators on the states of a spin-half particle. The Hopf fibration represents the 3-sphere as a circle fibration over the 2-sphere. The Bloch sphere describes the states of a two-state quantum system, through an analogy with the spin-half system. Mathematically, the $\IC^2/U(1)$ quotient is at the root of all of the above-mentioned topics. 

We record the form of several quantities, which appear in the expression for the kinetic energy, in terms of the $Q$ variables \begin{align}
 I &=  \frac{1}{M} \[ m_2\, m_3 (Q_0 - Q_1 ) + cyc. \]  \non 
 L_G &=  -\frac{\[ m_2\, m_3 \( (Q_1-Q_0-Q_4)\,  \frac{d}{dt} (Q_2-Q_3) - (Q_2-Q_3)\, \frac{d}{dt}(Q_1-Q_0-Q_4) \) + cyc.  \]}{2\, \sqrt{3} \,M\, (Q_0+Q_4)}  
\end{align}

\revision{
In planar motion, collinear configurations are codimension 1, and hence occur from time to time in generic trajectories, see e.g., \cite{Montgomery_2002}. Each time this occurs, the triangle changes its right/left handedness and the trajectory crosses the equator of the shape sphere. On the other hand, in 3d motion collinear configurations are codimension 2, and hence generic trajectories never cross the equator, rather they approach it from time to time only to be eventually repelled by the centrifugal force \eqref{def:cL2}. This raises a question regarding continuity of the limit of planar motion. A possible resolution is to identify $Q_4 \simeq -Q_4$. This would make the $Q$ variables invariant not only with respect to 2d rotations, but also with respect to 3d rotations.  
}

\subsection{Angular momentum variables}
\label{subsec:ang_mom}

In order to incorporate into the reduction the conservation of angular momenta, we transform from angular velocity variables into the conjugate momenta, which are none other than the angular momenta \be
	J_i = \frac{\del T}{\del \omega^i} \qquad i=1,2,3. 
	\label{def:J}
 \ee  
We have arrived at the final set of dynamical variables, namely \be
	\vec{J},\, \vec{g}
\label{dyn_var}
\ee
 where $\vec{g}$ is a location in geometry space, which usually would be represented by the spherical coordinates $r, \theta, \phi$.

Performing a Legendre transform over the Lagrangian \eqref{def:L_r}, we obtain \be
 	\cL_J := \cL -  \vec{\omega} \cdot \vec{J} = \cL_0 + \cL_1 + \cL_2 ~.
\label{L_J}
\ee	
$\cL_J$ is a function of the dynamical variables \eqref{dyn_var} (together with the generalized velocities associated with $\vec{g}$). The expression for $\cL_J$ is organized into three parts $\cL_0, \cL_1, \cL_2$ according to powers of $\vec{J}$, and their values are detailed below.

The transform that defined $\cL_J$ was taken only with respect to part of the velocity variables, and therefore, $\cL_J$  is a hybrid of Lagrangian and a Hamiltonian (sometimes known as Routhian). Later, we shall see that the equations of motion can be derived from $\cL_J$. It is useful to have a term that refers to any kind of such function, whether it is a Lagrangian, a Hamiltonian or hybrid. In mechanics, the standard potential function encodes the forces through derivatives. In thermodynamics, one uses one of several \emph{thermodynamic potentials}, such as the energy, the free-energy, the Gibbs free energy, all related among themselves by Legendre transforms. Similarly, we shall use the term \emph{motion potential} to refer to any function from which the equations of motion can be derived. From this perspective, the standard potential can be distinguished by the term \emph{force potential}.

The first term appeared already in (\ref{def:V},\ref{def:T_G}) and is repeated here for convenience. It depends only on the geometry space variables \be
	\cL_0 := \frac{1}{8\, I}\, \dot{I}^2  + \frac{3\, M_3}{8\, M} \frac{r^4}{I} \(\dot{\theta}^2 + \sin^2 \theta\, \dot{\phi}^2 \) +  \( \frac{G m_2\, m_3}{r_{23}} + cyc. \)
\label{def:cL0}
\ee
where \be
	I = \frac{1}{M} \( m_2\, m_3\, r_{23}^{~2} + cyc.\) 
\label{I_sides}
	\ee
and \begin{align}
	r_{23}^{~2} &= r^2 \(1 - \sin \theta\, \cos \phi \) \non
	r_{31}^{~2} &= r^2 \(1 - \sin \theta\, \cos (\phi-2 \pi/3) \) \non
	r_{12}^{~2} &= r^2 \(1 - \sin \theta\, \cos (\phi-4 \pi/3) \) 
\end{align}
The first two terms of $\cL_0$ specify the kinetic energy on geometry space -- in the radial and angular directions, respectively.  The last term is minus the potential, and for concreteness we present the case of a Newtonian gravitational potential.  

\vspace{0.5cm}

The $\cL_1$ term couples geometry space and the rotating body, and it is given by \be
	 \cL_1 := J_3 \frac{L_G}{I}
\label{def:cL1}
\ee
where $L_G$ is given by \be
	L_G = -\frac{r^2}{2\, M} \[ m_2\, m_3 \( \(1 - \cos \theta - \sin \theta \, \cos \phi \) \dot{\phi} - \sin \phi \, \dot{\theta} \) + cyc.  \] 
\ee
 $cyc.$ denotes $1 \to 2 \to 3 \to 1$ together with $\phi \to \phi - 2 \pi/3$. The expression for $L_G$ is given in the $\psi_+$ gauge \eqref{def:psi+}, and a gauge transformation with a gauge-function $\chi$ could shift it by $\Delta \cL_1 = - J_3\, \dot{\chi}$. In geometry space, $\cL_1$ is analogous in form to a coupling of the motion in $G$-space to a vector potential, as noticed by \cite{Lemaitre_1952}. This means that the motion in $G$-space experiences a velocity-dependent force, which is magnetic-like in form, and can be thought to originate from a Coriolis force.  
 
The magnetic-like vector is given by \be
  \vec{B} := 2 \frac{J_3}{I^{3/2}}\, r\, {\hat r}
 \label{B_field}
 \ee
 where ${\hat r}$ is a unit vector in the radial direction. It is obtained by dualizing the two-form associated with $\cL_1$ with respect to the kinetic metric \eqref{def:T_G}: $\vec{B} = *d\( J_3 \frac{L_G\, dt}{I} \) = 3 M_3\, J_3 \, r^4 \, / (2\, M\, I^2)\,  *\( d\theta \,  \sin \theta d\phi \)$ and using $*=1/\sqrt{g}\, \del^3 x = 8\, M\, I^{3/2}/(3 M_3 r^4 \sin \theta) \del_I \del_\theta \del_\phi$  we obtain $\vec{B}=4\, J_3 \del_I /\sqrt{I} = 2\, J_3 r \del_r /I^{3/2}$. The expression for the magnetic-like vector field is surprisingly simple. It is radial and inversely proportional to $r$, namely $\vec{B} \propto {\hat r}/r^2$ (since $I \propto r^2$).  It carries non-zero magnetic monopole charge, proportional to $J_3$ and located at $\vec{g}=0$. The monopole charge is a consequence of the intrinsic charge associated with the Hopf fibration, which is used to reduce over $\psi$ rotations. \revision{The magnetic-like vector field has appeared already, though in different form, in \cite{Lemaitre_1964} and as curvature terms in \cite{Moeckel_Mont_2013}.}

On the side of the rotating body, $\cL_1$ will be understood to generate a precession of $\vec{J}$ around the $z$-axis.

\vspace{0.5cm}

 The $\cL_2$ term also couples geometry space and the rotating body; only this term is quadratic in $\vec{J}$ and is given by \be 
	- \cL_2 := \frac{1}{2 I} J_3^{~2} + \frac{M}{2 M_3} \frac{1}{\Delta^2} \bar{I}^{\alpha\beta}\, J_\alpha J_\beta 
\label{def:cL2}
\ee 
 where $\Delta$ is twice the triangle's area, defined in \eqref{def:Delta}, and it can be expressed in $G$-variables through   \be
 \Delta^2 = \frac{3}{4} \, r^4\, \cos^2 \theta~,
 \ee
 and where $\bar{I}_{\alpha\beta}$ is proportional to the inverse of the $I_{\alpha\beta}$ matrix and is given by \be
	\bar{I}^{\alpha\beta} := \frac{1}{M} \( m_2\, m_3\,  \rho_{23}^\alpha\, \rho_{23}^\beta+ cyc. \)  
	\ee
where $\rho_{23}$ is given within the $\psi_+$ gauge by \be
	 \rho_{23} = i\, r\, \[  \cos \frac{\theta}{2}\, - e^{-i \phi}\, \sin \frac{\theta}{2}\, \] ~.
\ee

In geometry space, minus $\cL_2$ is interpreted as the \emph{centrifugal potential}.  It can be thought to generalize the familiar 2-body centrifugal potential \be
	  V_{\rm cent, 2\, body} = \frac{L^2}{2 \mu \, r^2}
\label{V_cent_2bd}
\ee
where $\vec{L}$ is the system's angular momentum, and $\mu$ is its reduced mass. Indeed, 2-body motion is necessarily planar and hence $J_1=J_2=0$. Moreover, $I=\mu r^2$, thereby the 3-body expression for the centrifugal potential reduces to that of the 2-body. 

As usual, a coupling in the motion potential implies several terms in the equations of motion. In this case, in addition to  a centrifugal force acting on the geometric variables $\vec{g}$, we shall see that $\cL_2$ also implies the Euler equations for the rotating body.

\presub {\bf Equations of motion}. We performed several natural changes of variables in the kinetic energy in order to re-formulate the problem, and thereby re-phrase the equations of motion. One possibility to achieve the equations of motion would be to express $\vec{\omega}$ in \eqref{T_in_G_var} in terms of a set of frame orientation angles, such as the Euler angles, which would be used as the fundamental dynamical variables. However, this procedure requires to make an arbitrary choice of the Euler-like angles and obscures the rotational symmetry. 

Alternatively, it would be nice to derive the equations of motion using $\vec{\omega}$ as fundamental velocities variables. However, the standard Euler-Lagrange equations would not produce the correct equations of motion, as can be seen for the example of the rigid body in appendix \ref{app:rigid}. Indeed, if we write $\omega^i = \beta^i_j\, \dot{q}^j$, where $q^i$ are generalized coordinates, then $\beta^i \equiv \beta^i_j\, dq^j$ are 1-forms over $G$-space, such that $d\beta^i \neq 0$ and hence $\beta^i$ cannot be expressed as a differential of any generalized coordinates. In other words, $\beta^i$ define a non-coordinate basis of differential forms, also known as a non-holonomic basis.

In order to derive the equations of motion from a Lagrangian expressed in terms of non-coordinate velocities, we rediscovered the appropriate generalization of the Euler-Lagrange equations. This partially known generalization was originally found by Poincare in 1901 \cite{Poincare1901} and it is described in Appendix \ref{app:rigid}. Here we only state the result for the case at hand.

In a Lagrangian formulation, one takes \be
 \cL(\vec{\omega},\vec{g}) = T(\vec{\omega},\vec{g}) - V(\vec{g})
\label{L_omega}
 \ee
 where $T(\vec{\omega},\vec{g})$ is given by \eqref{T_in_G_var} and $V(\vec{g})$ is an arbitrary potential.
The above-mentioned differential 1-forms have the following non-zero exterior differentials \be
 d\beta^i = \half \eps_{ijk}\, \beta^k \beta^j 
  \ee
 and hence the equations of motion read \bea
  \frac{d}{dt} \( \frac{\del \cL}{\del \omega^i} \) + \eps_{ijk}\, \omega^j \, \frac{\del \cL}{\del \omega^k}  &=& 0 \non
  \frac{d}{dt} \( \frac{\del \cL}{\del \dot{q}^s} \) 	&=&	\frac{\del \cL}{\del q^s}
 \eea
where $q^s=(r,\theta,\phi)$, the spherical coordinates in geometry space. The first line describes 3 generalized equations of motion which originate from variations with respect to $\omega$, while the second line describes 3 standard Euler-Lagrange equations of motion that originate from variation with respect to the $G$-space variables. 

In terms of the partial Legendre transform $\cL_J$ \eqref{L_J}, the equations of motion are given by \bea
 \frac{d}{dt}\, J_i	 &=& -\{ J_i, \cL_J \} \non
  \frac{d}{dt} \( \frac{\del \cL_J}{\del \dot{q}^s} \)	&=& \frac{\del \cL_J}{\del q^s} 
\label{eom_LJ}
\eea
where the Poisson brackets among the dynamic variables originate from the non-coordinate nature of the corresponding velocities according to the general rule \eqref{p_brack} and in our case are given by \be
 \{ J_i, J_j \} = -\eps_{ijk} \, J_k
\label{J_eq}
 \ee

After a solution $\vec{J}=\vec{J}(t),\, \vec{g}=\vec{g}(t)$ is found, one can further integrate to obtain the frame orientation angles.

Note that $\vec{J}^{\;2}$ is conserved as a result of (\ref{L_J},\ref{eom_LJ}). Hence, phase space is essentially 8 dimensional: 6 dimensions for $\vec{g}$ and its generalized velocities and 2 additional dimensions for $\vec{J}$ that lies on a sphere of fixed $\vec{J}^{\;2}$.

We comment that one could obtain a formulation where the vector potential is replaced by the gauge-invariant field-strength by performing a Legendre transform on the remaining velocity variables, and employing ``covariant momenta'' (in analogy with the covariant derivative).

\subsection{Discussion}
\label{subsec:discussion}

The motion potential $\cL_J=\cL_J(\vec{J},\vec{g})$ given by \eqref{L_J} defines a reformulation of the three-body problem, and it is a central result of this paper. The dynamic variables $\vec{J},\vec{g}$ are compatible with the conserved charges in such a way as to reduce the number of effective degrees of freedom. 

The employed dynamical variables decompose into two sets. The first set consists of the $\vec{J}$ variables that describe the angular momentum in the body frame. They are the momenta canonically conjugate to $\vec{\omega}$, the angular velocity vector in the body frame. Hence, $\vec{J}$ describes the rotational motion of the configuration triangle. The second set of variables consists of $\vec{g}$ or in spherical coordinates $(r,\theta,\phi)$. These describe the geometry of the triangle, namely its shape and size, see fig. \ref{fig1:shape_sphere}. The two sets are coupled as the geometry of the triangle determines its moment of inertia, which affects its rotation motion.  

\begin{figure}
\centering \noindent
\includegraphics[width=14cm]{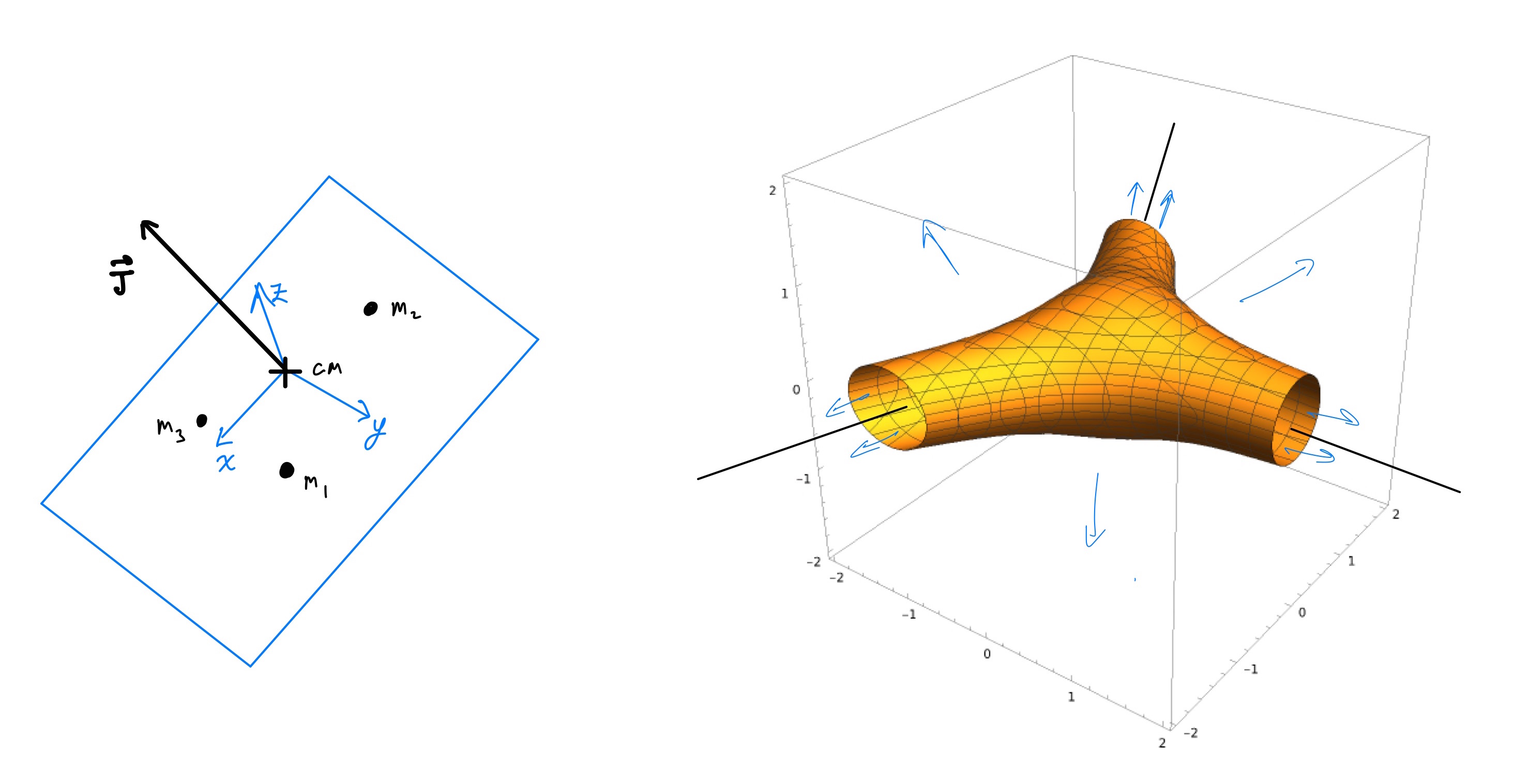} 
\caption[]{The natural dynamical reduction into orientation and geometry. Left: the three bodies define a triangle. Its dynamics is decomposed into its orientation and its geometry (shape + size). The dynamical variables for the orientation are taken to be the components of the total angular momentum within the rotating system. \\
Right: Mechanics in geometry space. Geometry space is a 3-dimensional space in which each point describes the shape and size of a triangle formed by the three bodies.  $r,\theta,\phi$ are considered to be spherical coordinates in this space. The (yellow) surface shaped like a pipe joint is a surface of constant potential (drawn for the equal mass case), so that motion is confined to be within it.  The three solid (black) lines are sources of an attractive potential which originates with the Newtonian gravitational potential. Finally, the (blue) arrows show the radial magnetic-like field, which originates from a Coriolis force.}
 \label{fig2:orientation-geometry} \end{figure}

The corresponding decomposition of the mechanics into a rotational motion and a motion in 3d geometry space is illustrated in figure \ref{fig2:orientation-geometry}. 

Let us look for the origin of the chaotic nature of the system in terms of these two components. Already in the limit of planar motion, the rotational component of the motion becomes trivial, yet the system is chaotic. Indeed, the motion in geometry space is non-integrable since it has 3 degrees of freedom and only a single conserved charge, namely the energy. Coupling to a rotating body cannot change this non-integrable nature. Moreover, given that the motion of a rotating body is integrable in the rigid body limit, \emph{geometry space can be considered to be the core chaotic component in the three-body system}. 

Various ingredients of this formulation have already appeared in the literature. The $r, \theta,\phi$ coordinates together with the vector potential appeared in \cite{Lemaitre_1952}. The kinetic term in shape space, which is conformal to the round sphere, appeared in \revision{\cite{Montgomery_2002,Moeckel_Mont_2013}} together with the origin of geometry space from a quotient of a 4d space. In hindsight, these elements partially appeared also in the Helium atom context:  the 3d geometry space and its conformal equivalence to the sphere in  \cite{Gronwall_1932}, and the 4d predecessor in \cite{Fock_1954}.

The current paper includes two main novelties. First, the definition of the complex position vector $\vec{w}$ that solves the center of mass constraint \eqref{def:w}. Secondly, the formulation in terms of $\vec{J}$, the angular momentum components in the rotating system, see subsection \ref{subsec:ang_mom}. In comparison to previous work, \cite{Moeckel_Mont_2013} was limited to planar motion, while we address the full 3d problem. $\vec{J}$ has not been used as dynamical variables before, and in particular neither in \cite{Lemaitre_1952} nor in \cite{Moeckel_Mont_2013}. While in hindsight, \cite{Tkachenko_1978} can be recognized to have some relation to a formulation in terms of $\vec{J}$ within the Helium atom context, it is at most partial and special due to the special values of the masses and the distinct quantum context.

Finally, the definition of $\vec{w}$ naturally explains certain features that were previously observed, yet deemed surprising.  \cite{Lemaitre_1952} introduces the spherical coordinates in geometry space without explanation. It starts with the equal mass case, where the correct definitions are easier to guess, before generalizing to the general case of unequal masses. \cite{Moeckel_Mont_2013}  arrives at these coordinates after a choice of coordinates on $\mathbb{CP}^1$ and comments that ``Remarkably, it turns out that if we put the binary collisions at the third roots of unity... then the equilateral points are automatically moved to the north and south poles.'' The definition of $\vec{w}$ provides the missing link and makes natural the transition to the geometry space coordinates and their properties. 

\section{Applications}
\label{sec:appl}

This section describes certain applications of the formulation in section \ref{sec:reduction}.

\subsection{Hill-like region in geometry space}

We have described a dynamical reduction onto geometry space, the space of all possible triangle geometries. One of its ingredients is a centrifugal potential, which implies an effective potential in geometry space. This allows to define the region of allowed motion in geometry space given the conserved charges. It is a generalization of the Hill region  from its original context within the hierarchical limit \cite{Hill1877,Hill1878}, into the full non-hierarchical range.

Let us rewrite the motion potential term \eqref{def:cL2} as a centrifugal potential (as remarked above \eqref{V_cent_2bd} ) \be 
 V_{\rm cent} = \half \(I^{-1}\)^{ij}\, J_i \, J_j ~.  
 \label{def:V_cent}
\ee

The principal moments of inertia are bounded from above by $I_a \le I, ~a=1,2,3$, where $I=I(\vec{g})$ is the moment around axis 3 that is perpendicular to the instantaneous plane and was given in (\ref{I-G},\ref{I_sides}) as a function of geometry space. 
This inequality holds since a three-body configuration is necessarily planar. It implies a lower bound on the centrifugal energy \be
  V_{\rm cent} \ge V_{\rm cent,min} := \frac{J^2}{2\, I} ~. 
\label{def:V_cent-min}
  \ee
 The bound is saturated when $\vec{J}$ is in the direction of axis 3, namely $J^2 = J_3^2$.
 
Given the conserved quantities $E,\, J^2$, the following inequality holds $E \ge V + V_{\rm cent} $, where the potential $V$ is given by \eqref{def:V}.  Combining with \eqref{def:V_cent-min}, the motion in geometry space is restricted to \be
 E \ge V\(\vec{g}\) + \frac{J^2}{2\, I\(\vec{g}\)} ~.
 \label{Hill_region}
 \ee
In fact, the inequality can be saturated -- this happens when both  $\dot{\vec{g}}=0$ and $J^2 = J_3^2$. Therefore, this relation defines the projection of the allowed phase space into geometry space, which deserves to be called a Hill-like region (or in short, a Hill region). 

The Hill region in geometry space is illustrated by the right hand side of figure \ref{fig2:orientation-geometry}, where the Hill region lies inside of the shown surface, which resembles a pipe joint connecting three pipes. In the figure, the masses are equal and  $J^2=0$. For $J^2 > 0$, a neighborhood of the origin, $\vec{g}=0$, is deleted due to the centrifugal term. As $J^2$ is increased, the Hill region shrinks and can be seen to undergo a topology change.

\revision{
A generalization of Hill's region to general masses appeared in \cite{Marchal_Saari_1975},\footnote{
I thank D. Fabrycky for this reference and the following one.}  and their eq. (3.11) [$2 \rho/\nu \ge \rho/a + p/\rho$ in their notation] is equivalent to (\ref{Hill_region}). See also  \cite{Marchal_Bozis_1982} and references therein. 
 In this work, Hill's region is defined in geometry space. 
An analogous discussion of Hill's region for the planar problem can be found in \cite{Moeckel_1988}. The dynamical implications found there are hereby generalized to the 3d system. 
}
\subsection{The total phase-volume}

The dynamical reduction was applied to the evaluation of the regularized phase volume of the three-body system.

The phase-volume is of intrinsic interest, and its regularized version is an ingredient of the flux-based theory \cite{flux_based}. It is defined through a multi-dimensional integral over all of phase space. The dynamical reduction motivates changes of variables that allow to perform some of these integrations analytically, thereby enabling a full evaluation, as shown in \cite{sigma}. 

In fact, phase volume evaluation serves not only as an application of the dynamical reduction, but also as a signpost on the road to it, as described in the introduction. Therefore, we can consider the two topics to have co-evolved in symbiosis.

\subsection{Planar motion}

This subsection describes an application of the current formulation to planar three-body motion, and in particular to uniformly rotating configurations.

If the initial velocities are within the three-body plane, then the motion remains planar throughout.  For planar motion, $J_1, J_2$ vanish identically, while $J_3$ is conserved. Hence, the rotating body motion is trivial, and $\cL_2$ specializes to \be 
\cL_{2,2d} = - \frac{1}{2 I} J_3^{~2}
 \ee
where $\cL_0, \cL_1$ are defined in (\ref{def:cL0},\ref{def:cL1}).

\presub {\bf Equilibria}. Equilibria of the reduced planar motion describe rigidly rotating solutions. 

The effective potential in geometry space $V_{eff}^G = V_{eff}^G (r,\theta,\phi)$ is given by \be
 	V_{eff}^G = V + \frac{J_3^{~2}}{2\, I}
 \ee
where the second term is the centrifugal potential.

 If $V$ is a power-law in the inter-body distance, one can solve the equilibrium potential explicitly with respect to $r$. In the remainder of this section, we assume the Newtonian gravitational potential. Solving the equation $0=\del V_{eff}^G/\del r$ for $r$ and substituting back into $V_{eff}^G$ one gets a potential that may be called the effective potential in shape space  $V_{eff}^S = V_{eff}^S (\theta,\phi)$. It is given by \be
	V_{eff}^S  = -\frac{1}{2\, J_3^{~2}} I\, V^2
\ee	
Note that $I\, V^2$ is indeed $r$-independent, on account of the $r$-scaling properties of $I$ and $V$.

To proceed towards the equilibria, it remains to differentiate $V_{eff}^S$ with respect to $\theta, \phi$. It turns out to be convenient to use the side-lengths, $r_{12}, r_{23}, r_{31}$, rather than $\theta, \phi$. We require \begin{align}
 	0 &= \frac{\del}{\del r_{23}} \log V_{eff}^S =  \frac{\del}{\del r_{23}} \( \log I + 2 \log (-V) \) = \non
		&= \frac{2\, m_2\, m_3\, r_{23}}{M\, I} -  \frac{2\, G\, m_2\, m_3}{(-V)\, r_{23}^{~2}}
\end{align}
 where we have used the expressions for $V,\, I$ from (\ref{def:V}, \ref{I_sides})	.  Solving for $r_{23}$ the factors of $m_2\, m_3$ cancel out on account of the equality of gravitational and inertial masses, and one finds that the side lengths are all equal \be
  r_{12}^{~3} = r_{23}^{~3} = r_{31}^{~3} =  \frac{M\, J_3^2}{G\, M_2^2}  ~.
  \ee
These are \emph{the equilateral solutions found by Lagrange} \cite{Lagrange_1772}. Thus, we presented \emph{a derivation of them through the reduced Lagrangian in geometry space}. 

Note that the angular velocity associated with an equilateral with sides $a$ is given by \be
 \Omega^2 \equiv \( \frac{J_3}{I} \)^2 = \frac{GM}{a^3}
 \label{Omega}
 \ee
 which is consistent with Kepler's third law and the fact that within the equilateral orbit each body moves as though attracted toward the center of mass (for this reason, these configurations are known as central configurations).  
 
\presub We add some comments.

Collinear solutions.  $V_{eff}^S$ is invariant under a space reflection, which when translated to geometry space becomes a reflection through the equator. Hence, the gradient of $V_{eff}^S$ must lie within the equator. By restricting $V_{eff}^S$ to the equator one finds the three collinear solutions found by Euler \cite{Euler_1767}. These solutions were not found above while differentiating with respect to the $r_{ab}$ variables, since the transformation between them and the $r,\theta,\phi$ variables is singular at the equator. 

Type of equilibrium. By noting the behavior of $V_{eff}^S$ near the equator, one concludes that the extremum associated with equilateral motion is in fact a maximum. 

Stability. The condition for the stability of the equilateral solutions is  $M_2/M^2 \le 1/27$. It was found by Gascheau (1843) \cite{Gascheau1843}, see also \cite{Routh1875} since I was not able to locate the former reference. 

More generally, central configurations yield not only rigidly rotating solutions but also rotating--rescaling elliptic solutions, where the masses form an equilateral triangle and each mass moves on a Keplerian ellipse.  In geometry space, these solutions correspond to radial motion along the polar directions, which can be seen to be a consistent truncation of the equations of motion. 
 This validates the term $\dot{I}^2/(8\, I) \subset T_G$ \eqref{def:T_G}.  The stability of the elliptic equilateral solutions was found in \cite{Danby1964,RobertsG2002,Sicardy2010,Martinez_Sama_Simo_2006}. 

The current formulation reduces the stability analysis. In particular, the stability of the circular equilateral solutions is reduced to stability around a static solution. It would be interesting to revisit the above-mentioned stability criteria. 

\subsection{Economizing simulation}

\revision{
Clearly, a formulation of a dynamical reduction has implications for a reduction of computation time during simulations (in addition to implications to theory).

The standard Newtonian formulation involves 9 second order differential equations (or equivalently, 9 degrees of freedom).

Using the $j$-complex position vector $\vec{w}$ \eqref{def:w} as a dynamical variable for simulations reduces the equation set to 6 second order equations (6 degrees of freedom) and guarantees the conservation of the center of mass during the simulation. In other words, it avoids the cost of an unnecessary simulation of the motion of the center of mass. 
 
In addition, transforming the equations into the geometry variables $\vec{g} \leftrightarrow (r, \, \theta, \, \phi)$ \eqref{def:spher} and the angular momentum variables in the body frame $\vec{J}$ \eqref{def:J}, reduces the equations to 3 second order equations for $\vec{g}$, plus 3 first order equations for $\vec{J}$ (essentially, Euler's equations), see \eqref{eom_LJ}, plus 3 integrations of Euler-like orientation angles, which do not affect the previous equations. In fact, $\vec{J}$ is constrained to move on the surface of a sphere. Hence, by transforming to coordinates on this sphere, the $\vec{J}$ equations would reduce to 2 first-order equations, which guarantee the conservation of ${\vec J}^2$.

It would be interesting to implement these equation sets on a computer and to measure or quantify the resulting reduction in simulation time. 
}

\section{Four-body problem}
\label{sec:4body}

In the three-body problem, we introduced symmetric vector coordinates in the center of mass frame, inspired by Lagrange's solution to the cubic \eqref{def:w}. The quartic equation also has a general solution, which suggests a generalization to the four-body problem.

Denoting the masses by $m_a, ~a=1,2,3,4$ and the positions by  $\vec{r}_a$ we define \bea
 \vec{s}_1 &=& \vec{r}_1 - \vec{r}_2 - \vec{r}_3 + \vec{r}_4 \non
 \vec{s}_2 &=& -\vec{r}_1 + \vec{r}_2 - \vec{r}_3 + \vec{r}_4 \non
 \vec{s}_3 &=& -\vec{r}_1 - \vec{r}_2 + \vec{r}_3 + \vec{r}_4 
 \eea
 These variables are translation-invariant vectors. They contain 9 degrees of freedom, which are necessary to cover the configuration space at the center of mass frame. As the labels $1,2,3,4$ are permuted, the $\vec{s}$ vectors transform nicely -- they permute among themselves and/or change signs. 

In order to proceed and decompose the variables into orientation and geometrical variables, the invariants are the 6 scalar products $Q_{rt}= \vec{s}_r \cdot \vec{s}_t, ~ r,t=1,2,3$ and the gauge-field and associated field-strength become $SO(3)$-valued, and so, non-Abelian. 

\section{Epilogue}
\label{sec:epilogue}

\revision{
\cite{Lagrange_1772} reduced the formulation of the three-body problem based on mutual distance variables. \cite{Lemaitre_1952} introduced triangle geometry variables that make the collinear configurations regular. 

This paper incorporates two main novelties into the formulation. First, it defines the complex position vector $\vec{w}$ \eqref{def:w} that provides a missing link towards the geometry space variables and a motivation for them. Secondly, it introduces into the formulation the angular momenta in the rotating frame $\vec{J}$. Several applications were discussed.
}

\subsection*{Acknowledgments}

I thank Yogesh Dandekar, Lior Lederer and Subhajit Mazumdar for collaboration on a related project. This research, as well as \cite{flux_based}, was performed during the times of the COVID-19 pandemic and benefitted from the associated isolation. Part of this research was supported by the Israel Science Foundation (grant no. 1345/21).

I dedicate this work to the memory of Ami Nathan, 15.6.1934--24.10.2020, my father in law and a dedicated family man.

\subsection*{Data availability}

No new data were generated or analysed in support of this research.

\appendix

\section{Quotes from previous works}
\label{app:quotes}

In this section, we collect some selected quotes from previous work for the curiosity of the interested reader.

In the Principia \cite{Principia} Newton states his aim for studying the Moon--Earth--Sun three-body system: ``By these computations of the lunar motions I was desirous of showing that by the theory of gravity the motions of the moon could be calculated from their physical causes. The calculus of this motion is difficult ...'' (in the Scholium following Proposition XXXV, Book III, this translation appears in \cite{Principia_Chandra} p. 419).

Sir David Brewster's 1855 biography of Newton \cite{Brewster_1855} elaborates on the above-mentioned difficulty "... we find him [Newton] occupied with the difficult and profound subject of the lunar irregularities. He had resumed this inquiry in 1692, and it was probably from the intense application of his mental powers which that subject demanded, that he was deprived of his appetite and sleep during that and the subsequent year. When Mr. Machin long afterwards was complimenting him upon his successful treatment of it, Sir Isaac told him that his head had never ached but when he was studying that subject; and Dr. Halley told Conduitt that he often pressed him to complete his theory of the moon, and that he always replied that it made his head ache, and kept him awake so often, that he would think of it no more.'' 

This part of the Principia is arguably the first appearance of perturbation theory in physics (since perturbation theory aims to determine a series of consecutive corrections in some small parameters, it rests essentially on the differential calculus invented by Newton, and moreover, if we accept the view that the Principia can be marked as the starting point of physics, then there could be no earlier appearance of perturbation theory in physics). Anyone who has performed the meticulous accounting required by perturbation theory, will identify with Newton's headache. 

The last quote regarding Newton's work is Chandrasekhar's 1995 extraordinary evaluation ``In summary, one may in truth say that there is hardly anything in any modern textbook
on celestial mechanics 
 that one cannot find in the propositions that we have enumerated, and indeed with deeper understanding.'' \cite{Principia_Chandra} p.419.

In the opening to his work on the solution to algebraic equations \cite{Lagrange_1771} (in the fourth paragraph),  Lagrange writes ``Je me propose dans ce M\'emoire d'examiner les diff\'erentes m\'ethodes que l'on a trouv\'ees jusqu'\`a pr\'esent pour la r\'esolution alg\'ebrique des \'equations, de les r\'eduire \`a des principes g\'en\'eraux, et de faire voir {\it \`a priori} pourquoi ces m\'ethodes r\'eussissent pour le troisi\`eme et le quatri\`eme degr\'e, et sont en d\'efaut pour les degr\'es ult\'erieurs.''  [In free translation:  In this memoir, I propose to examine the different methods that have been found so far for the algebraic solution of equations, to reduce them to general principles, and to show {\it a priori} why these methods succeed for the third and the fourth degree, and are in default for the higher degrees.]

On the following year, Lagrange opens the essay on the three-body problem \cite{Lagrange_1772}, which won a prize from the French royal academy of sciences, with ``Ces recherches renferment une M\'ethode pour r\'esoudre le Probl\`eme des trois Corps, diff\'erente de toutes celles qui ont \'et\'e donn\'ees jusqu'\`a pr\'esent. Elle consiste \`a n'employer dans la d\'etermination de l'orbite de chaque Corps d'autres \'el\'ements que les distances entre les trois Corps, c'est-\`a-dire, le triangle form\'e par ces Corps \`a chaque instant.'' [In free translation: This research contains a method for solving the problem of the three bodies, different from all those which have been given until now. It consists in not employing in the determination of the orbit of each body elements other than the distances between the three bodies, that is to say, the triangle formed by these bodies at each instant.]

Jacobi opens the paper on the elimination of the nodes \cite{Jacobi_1842} with ``Les illustres g\'eom\`etres du si\`ecle pass\'e, en traitant le probl\`eme des trois corps,...'' [in free translation: The illustrious geometers of the past century, in dealing with the problem of three bodies,...] and ends it with ``Voil\`a donc le probl\`eme de trois corps r\'eduit \`a l'int\'egration des six \'equations (I \`a VI) et \`a une quadrature. Les six \'equations diff\'erentielles (I \`a VI) sont toutes du premier degr\'e, hors une seule qui est du second, et it n'y entre aucune trace des noeuds.'' [In free translation: Here, then, the three-body problem is reduced to the integration of six equations (I to VI) and a quadrature. The six differential equations (I to VI) are all of the first degree, apart from only one which is of the second, and there is no trace of the nodes.]

In  Murnaghan's 1936 paper \cite{Murnaghan_1936}, the title reads ``A symmetric reduction of the planar three-body problem''.

The abstract to Lema\^itre's 1952 paper \cite{Lemaitre_1952} reads ``Le probl\`eme des trois corps dans le cas plan et pour des masse \'egales est equivalent au mouvement d'un point sous l'action d'une fonction des forces et d'un potentiel vecteur. On montre comment dans le cas g\'en\'eral l'asym\'etrie introduite par l'in\'egalit\'e des masses modifie l'hamiltonien.'' [In free translation:  The three-body problem in the planar case and for equal masses is equivalent to the motion of a point under the action of a function of forces and a vector potential. We show how in the general case, the asymmetry introduced by the mass inequality modifies the Hamiltonian.]

\revision{
In the introduction to \cite{Lemaitre_1964} Lema\^itre writes ``I introduce relative axes which are the principal axes of inertia of the system of the three masses''. Later in the same section ``In this former work the author had been able to use the principal axis only in the case where the three masses were equal. Many useless complications result from this shortcoming.''
}

Finally, Mongomery and Moeckel write in the abstract of their 2013 paper \cite{Moeckel_Mont_2013} ``We carry out a sequence of coordinate changes for the planar three-body problem, which successively eliminate the translation and rotation symmetries... Using size and shape coordinates facilitates the reduction by rotations ... while emphasizing the role of the shape sphere. By using homogeneous coordinates to describe Hamiltonian systems whose configurations spaces are spheres or projective spaces, we are able to take a modern, global approach to these familiar problems.''

\section{Lagrange's solution to the cubic}
\label{app:cubic}

In this appendix, we provide a bit more information on Lagrange's solution to the cubic.

A cubic is an equation of the form \be
 a\, x^3 + b\, x^2 + c\, x + d =0 ~.
 \ee

Denoting the roots by $x_1,\, x_2,\, x_3$ Lagrange defines \cite{Lagrange_1771} \bea
s_1 &:=& x_1 + \eta\, x_2 + \etab\,  x_3 \non
s_2 &:=& x_1 + \etab\, x_2 + \eta \, x_3 
\label{def:svar}
\eea
where $\eta$ is the cubic root of unity. \revision{$s_1, s_2$ are  known as Lagrange resolvents.}

Under cyclic permutations of $x_a, a=1,2,3$, the variables $s_1, s_2$ transform by a third of a revolution in phase. Hence, $s_1^{~3}$ and $s_2^{~3}$ are invariant under cyclic permutations. Moreover, any exchange permutation of $x_a, a=1,2,3$ exchanges $s_1^{~3}$ and $s_2^{~3}$. Hence, $S:=s_1^{~3} + s_2^{~3}$ is a fully symmetric function of $x_a, a=1,2,3$ and so is $P:= s_1\, s_2$. Therefore, by the fundamental theorem of symmetric polynomials, both $S,\, P$ can be expressed in terms of $b/a,\, c/a,\, d/a$. Now $z=s_1^{~3}, s_2^{~3}$ are the roots of the quadratic \be
	z^2 - S\, z + P^3 =0	~.
\ee
After solving this quadratic, one takes cubic roots to obtain $s_1, s_2$. Finally, \eqref{def:svar} together with $-b/a = s_0 := x_1+ x_2 + x_3 =$ is a system of linear equations which can be inverted to produce the three roots $x_a, a=1,2,3$. These transformations between the $x$ and $s$ variables are known as a discrete Fourier transform.

In this way, Lagrange's method relies on the symmetry properties of expressions in the roots in order to solve the cubic. This is a crucial insight on the road to Galois theory.

The $s_1, s_2$ variables are invariant under a shift in $x$ and have nice transformation properties under permutations. In this way, they inspired the definition of $\vec{w}$ \eqref{def:w}. 

\section{Bi-complex numbers and the isotropic oscillator}
\label{app:bicomplex}

A bi-complex number $w$ is an expression of form \be
	w= a + b\, i + c\, j + d \, i j
\ee 
where $a, b, c, d$ are real numbers.

Bi-complex numbers can be added and multiplied using the relations \bea
	 -1&=& i^2 = j^2 \non
	  i\, j &=& j\, i 
\eea
These relations define an algebraic structure known as a commutative ring. Note that bi-complex numbers are distinct from Quaternions: while both are dimension 4 over the Reals, the latter is non-commutative, e.g., $i\, j = - j\, i =k$.

Not all non-zero bi-complex numbers have an inverse. This is true of zero-divisors, namely non-zero elements $u, v$ such that $u \cdot  v =0$. The zero-divisors can be generated from the elementary zero divisors \be
	  R :=  1 + i\, j  \qquad L :=  1 - i\, j 
\ee
The notation $R,L$ stands for right and left: for the isotropic oscillator these states correspond to right/left circular motion, while for the three-body problem they correspond to right/left equilateral triangles.

Indeed, $R \cdot L =0$. Moreover, all other zero divisors can be generated from these through multiplication by a bi-complex number. 
$R, L$ satisfy the relations \bea
	 i\, R &=& - j \, R \non
	 i\, L &=& + j \, L \non
\eea
We note that after multiplication $R,L$ obtain the following simple form: $i R=i-j, ~ i L=i+j$.

\presub {\bf The isotropic oscillator}. The isotropic oscillator is the mechanical system consisting of a point mass $m$, free to move in the plane under the influence of a harmonic restoring force $\vec{F}=-m \omega^2\, \vec{r}$. It can be defined through the Hamiltonian \be
	 H = \frac{1}{2 m} \vec{p}^{\;2} + \half m\, \omega^2\, \vec{r}^{\;2} ~.
 \ee

One complexifies phase space through the definition of a complex vector  \be
	 \vec{a} := \frac{1}{\sqrt{2}} \( \sqrt{m \omega}\, \vec{r} + \frac{i}{ \sqrt{m \omega}}\, \vec{p} \) 
 \ee
In terms of these variables, one has \be
	 H = \omega\, \vec{a} \cdot  \vec{a}^* 
 \ee
 where $*$ denotes complex conjugation,  and the Poisson brackets \be
	 \{ a_\alpha, a^*_\beta \} = -i \, \delta_{\alpha\beta} 
 \ee
  --- all other Poisson brackets among the complex variables vanish. 
  
It is natural to further complexify the plane. Since plane rotations commute with phase space rotations, we use a different imaginary unit $j$ which commutes with $i$ and define a bi-complex number through \be
	 \vec{a} \to a := \vec{a} \cdot \( \hat{x} +j \, \hat{y} \)
  \ee

Since time evolution is given by \be
\vec{a}(t) = \exp \( i \omega t\) \, \vec{a}(0) ~, \ee 
the conserved charges can be expressed in terms of the Stokes parameters $S_0, S_1, S_2, S_3$ defined by 
\bea 
	S_0 	+ i\, j\, S_3	&:=& a \cdot \bar{a}^* \non
	S_1 + j\, S_2 	&=& a \cdot a^* 	~,
\eea
where a bar denotes conjugation with respect to $j$, and a $*$ continues to denote conjugation with respect to $i$. These expressions are equivalent to $S_0=\vec{a} \cdot \vec{a}^*, ~ S_i=\sigma^{i-1}_{\alpha \beta}\, a^*_\alpha\, a_\beta, ~i=1,2,3, ~\alpha,\beta=1,2$ where $\sigma^i, i=1,2,3$ are the Pauli matrices. If $(a_1,a_2)$ undergo a unitary transformation, $\vec{S} \equiv (S_1,S_2,S_3)$ undergoes an orthogonal transformation.  

The Stokes parameters satisfy the relation \be
	S_0^{\;2} = \vec{S}^2 ~.
\ee
In addition, their Poisson brackets are given by \bea
 	\{ S_i, S_j \} &=& \eps_{ijk}\, S_k \non
	 \{ S_i, S_0 \} &=& 0~.
 \eea


\section{Equations of motion in terms of non-coordinate velocities and the rotating rigid body}
\label{app:rigid}

Consider an $n$-dimensional configuration space $\cM$, described by some generalized coordinates $\( q^i \)_{i=1}^n$, as well as non-coordinate velocities (also known as quasi velocities \cite{Arnold1988}) \be
 v^a = \beta^a_i\, \dot{q}^i, ~ a=1,\dots, n
\label{def:v_a}
 \ee
where $\beta^a_i = \beta^a_i ( q^j )$ defines a changes of basis from the coordinate velocities $\dot{q}^i$ to $v^a$. As an example, we shall see below that the angular velocity components of a rotating rigid body are non-coordinate velocities.

Equivalently, the velocities define a basis of 1-forms over $\cM$ \be 
	\beta^a := \beta^a_i\, dq^i
\ee 

A basis of 1-forms is called a coordinate basis, or a holonomic basis, if the 1-forms can all be expressed as differentials of some functions. The non-coordinate nature of a basis is captured by the torsion coefficients  $C^c_{~ab}= C^c_{~ab} (q^i)$ defined by \be
 d\beta^c = \half\,  C^c_{~ab}\, \beta^b\, \beta^a ~.
\label{def:torsion}
 \ee
The dual basis of vector fields $e_a$ is defined by $\beta^a \cdot e_b = \delta ^a_b$. In components $e_b=e_b^i \del_i$ and $e_b^i$ is the inverse matrix for $\beta^a_i$. In terms of $e_a$, the structure constants can be defined equivalently through the Lie brackets $[e_a, e_b]= C^c_{~ab}\, e_c$.

Suppose the Lagrangian is given in terms of the generalized coordinates and the velocities $v^a$ \be
 \cL = \cL \( q^i, \, v^a \) ~. 
\ee
The generalization of the Euler-Lagrange equations of motion to a non-coordinate basis are \be
	\frac{d}{dt} \( \frac{\del \cL}{\del v^a} \) +  C^c_{ab}\, v^b \frac{\del \cL}{\del v^c} = \del_a \cL 
\label{mod_eom}
\ee
where $\del_a \equiv e^i_a\, \del_i$ is a derivative with respect to the vector field $e_a$. This generalization was found by Poincar\'e (1901) \cite{Poincare1901}, see also the books \cite{Arnold1988,Holm2011}, and the paper \cite{HolmMarsdenRatiu1998}.

For coordinate velocities  $C^c_{ab}=0$ and the equations reduce to the Euler-Lagrange equations.

\presub{\bf Derivation}. The derivation of the Euler-Lagrange equations relies on the commutation of the variation and the time derivative. For non-coordinate velocities, this no longer holds, and a correction term arises. Indeed \bea
 \delta v^a &\equiv& \delta \( \beta^a_i\, \frac{d}{dt} q^i \) = \frac{d}{dt} \(  \beta^a_i\, \delta q^i \) + 2\, \(d\beta^a\)_{ij} \delta q^i\, \dot{q}^j = \non
  &=& \frac{d}{dt} \(  \beta^a_i\, \delta q^i \) + C^a_{~bc}\, \delta q^c\, \dot{q}^b
\label{delta_v}
\eea
 where the last equality uses \eqref{def:torsion}.
 
Now the variation of the action can be performed \bea
 \delta S &=& \int \[ \frac{\del \cL}{\del q^i}\, \delta q^i + \frac{\del \cL}{\del v^a}\, \delta v^a \] = \non
 		&=&  \int \[ \frac{\del \cL}{\del q^i}\, \delta q^i + \frac{\del \cL}{\del v^a}\,  \frac{d}{dt} \(  \beta^a_i\, \delta q^i \) + \frac{\del \cL}{\del v^c}\,  C^c_{~ba}\, \delta q^a\, \dot{q}^b \] = \non
		&=& \int \delta q^a \[  \del_a \cL - \frac{d}{dt}\( \frac{\del \cL}{\del v^a} \) - C^c_{~ab}\,  v^b\, \frac{\del \cL}{\del v^c} \]
 \eea 
where in passing to the second line, we have used \eqref{delta_v}, and in passing to the third, we have used $q^i = e^i_a\, q^a$ and $C^c_{~ba} = - C^c_{~ab}$. The last line implies the modified equations of motion \eqref{mod_eom} --- QED.

\presub{\bf Momentum space}. This procedure can be generalized so that it applies to Hamiltonians as well. Define the momenta conjugate to $v^a$ \be
 p_a = \frac{\del \cL}{\del v^a} 
 \ee
These momenta can be expressed in terms of momenta conjugate to $\dot{q}^i$ through $p_a = e^i_a\, p_i$, where $e^i_a$ were defined below \eqref{def:torsion}. This implies non-zero Poisson brackets for $p_a$  \be
\{ p_a, p_b \}= -C^c_{~ab}\, p_c  
\label{p_brack}
\ee
From here one proceeds as usual to define the Hamiltonian through the Legendre transform, and to obtain the equations of motion for any dynamical variable $A=A(q^i,p_a)$, through $dA/dt = \{ A, \cH \}$. \revision{The Hamiltonian formulation in a non-coordinate basis appeared already in \cite{HolmMarsdenRatiu1998}. 
} 

\presub {\bf Lagrangian for Euler's rigid body equations}. Let us see how this enables a Lagrangian formulation for Euler's equations for a rotating rigid body.

A freely rotating rigid body is described by Euler's equations \be
0 =  \left. \frac{d}{dt} \right|_{inert} \vec{S} \equiv \dot{\vec{S}} + \vec{\omega} \times \vec{S} ~.
\ee Equivalently, in coordinate notation \be
 0= \left. \frac{d}{dt} \right|_{inert} I_1\, \omega_1 :=  I_1 \dot{\omega}_1 + (I_3 - I_2)\, \omega_2\, \omega_3
 \label{Euler_eom}
 \ee
 and similarly, two other equations by cyclic permutations. $\vec{\omega}$ is the angular velocity components in the frame defined by the body's principal axes, $I_i, ~i=1,2,3$ are the principal moments of inertia, and $\left. \frac{d}{dt} \right|_{inert}$ denotes a time derivative in the inertial frame.
 
 The kinetic energy of the body is given by  $T = \half\, I_{ij}\,  \omega^i\,  \omega^j$. A free body has no potential energy and hence  the Lagrangian is defined to be  \be
 \cL := T =  \half\, I_{ij}\,  \omega^i\,  \omega^j ~.
 \ee
 
If one applies the standard Euler-Lagrange equations carelessly, one finds $(d/dt) \(\del \cL/\del \omega^i\) = 0$. This is different from the Euler equations of motion \eqref{Euler_eom}. However, we notice that $\omega^i$ are non-coordinate velocities with torsion coefficients \be
 C^i_{jk} = \eps_{ijk}
 \label{rigid_C}
 \ee
 where $\eps_{ijk}$ is the Levi-Civita tensor. This can be shown by expressing $\omega^i$ in terms of Euler angles,\footnote{
 E.g., \bea
  \omega^x &=& -\sin \theta_n \, \cos \psi_n\, \dot{\phi}_n + \sin \psi_n\, \dot{\theta}_n \non 
  \omega^y &=&  \sin \theta_n \, \sin \psi_n\, \dot{\phi}_n + \cos \psi_n \, \dot{\theta}_n \non
  \omega^z &=& \cos \theta_n \, \dot{\phi}_n + \dot{\psi}_n
 \eea
 where $\theta_n,\, \phi_n,\, \psi_n$ are Euler-like angles. These expressions are of the form \eqref{def:v_a}.}
 and we note that the coefficients are constant over $\cM$ and coincide with the structure constants of $SO(3)$. The modified equations of motion \eqref{mod_eom} read \be
  0 =   \dot{S^i} +\eps^k_{ij} \omega^j S_k = \dot{S^i} + (\vec{\omega} \times \vec{S})^i
 \ee
 which now fully agree with Euler's equations. 
 
 We can now proceed to a Hamiltonian formulation. The conjugate momenta \be
 S_i = \frac{\del \cL}{\del \omega^i} = I_{ij} \, \omega^j
 \ee
 are the coordinates of the spin (angular momentum) vector in body coordinates.
 By (\ref{p_brack}, \ref{rigid_C}) their Poisson brackets are given by \be
 \{ S_i,S_j \} = -\eps_{ijk} \, S_k
 \ee
 
 The Hamiltonian is given by \be
  \cH =  \omega^i\, S_i - \cL = \half \(I^{-1}\)^{ij}\, S_i \, S_j
  \ee
  where $I^{-1}$ is in the inverse matrix for the moment of inertia tensor. Finally, the equations of motion are given by \be
  \frac{d}{dt}\, S^i = \{ S^i, \cH \} = \{ S^i, S^j \} \frac{\del \cH}{\del S^j} = - \eps_{ijk} \, \omega^j \, S^k
  \ee
  which reproduce the Euler equations. 
    
 \newpage
\bibliographystyle{unsrt}

\end{document}